\documentclass[pra,twocolumn,superscriptaddress]{revtex4-1}

\usepackage{amsmath}    
\usepackage{graphicx}   
\usepackage{hhline}
\usepackage{verbatim}   
\usepackage{color}      
\usepackage{hyperref}   
\usepackage{longtable}
\usepackage{float}
\usepackage[caption=false]{subfig}
\usepackage{grffile}
\usepackage{changepage}
\usepackage{amssymb}
\usepackage{siunitx}
\begin{document}

\title{Two-Photon Spectroscopy of the NaLi Triplet Ground State}
\author{Timur M. Rvachov}
\affiliation{Research Laboratory of Electronics, MIT-Harvard Center for Ultracold Atoms,
Department of Physics, Massachusetts Institute of Technology, Cambridge, Massachusetts 02139, USA}

\author{Hyungmok Son}
\affiliation{Research Laboratory of Electronics, MIT-Harvard Center for Ultracold Atoms,
Department of Physics, Massachusetts Institute of Technology, Cambridge, Massachusetts 02139, USA}
\affiliation{Department of Physics, Harvard University, Cambridge, Massachusetts 02138, USA}

\author{Juliana J. Park}
\affiliation{Research Laboratory of Electronics, MIT-Harvard Center for Ultracold Atoms,
Department of Physics, Massachusetts Institute of Technology, Cambridge, Massachusetts 02139, USA}

\author{Sepehr Ebadi}
\affiliation{Department of Physics, Harvard University, Cambridge, Massachusetts 02138, USA}

\author{Martin W. Zwierlein}
\affiliation{Research Laboratory of Electronics, MIT-Harvard Center for Ultracold Atoms,
Department of Physics, Massachusetts Institute of Technology, Cambridge, Massachusetts 02139, USA}

\author{Wolfgang Ketterle}
\affiliation{Research Laboratory of Electronics, MIT-Harvard Center for Ultracold Atoms,
Department of Physics, Massachusetts Institute of Technology, Cambridge, Massachusetts 02139, USA}

\author{Alan O. Jamison}
\affiliation{Research Laboratory of Electronics, MIT-Harvard Center for Ultracold Atoms,
Department of Physics, Massachusetts Institute of Technology, Cambridge, Massachusetts 02139, USA}

\date{\today}

\begin{abstract}

We employ two-photon spectroscopy to study the vibrational states of the triplet ground state potential ($a^3\Sigma^+$) of the $^{23}$Na$^{6}$Li molecule. Pairs of Na and Li atoms in an ultracold mixture are photoassociated into an excited triplet molecular state, which in turn is coupled to vibrational states of the triplet ground potential. Vibrational state binding energies, line strengths, and potential fitting parameters for the triplet ground $a^3\Sigma^+$ potential are reported. We also observe rotational splitting in the lowest vibrational state.

\end{abstract}

\maketitle

Ultracold gases of dipolar molecules are promising for a host of new scientific directions \cite{frontiers_ye}.  In the area of quantum simulation, they will enable many-body physics with long-range and anisotropic interactions \cite{review_ye_2009}. In precision measurements of fundamental constants, they may improve the precision in searches for the electric dipole moment of an electron \cite{edm_acme,edm_loh,edm_hinds}. For quantum chemistry, they provide a system with exceptional quantum state control for studies of chemical processes \cite{chemistry_review_krems_2008}. In contrast to many atomic species, ground-state molecules have not yet been cooled to quantum degeneracy, due to more complicated level schemes and less favorable collisional properties. Current work is focused on two approaches:  (i) the direct laser cooling of molecules pre-cooled by a buffer-gas source \cite{mot_demille_nature} and (ii) the coherent formation of molecules from ultracold mixtures of their constituent atoms \cite{cs2_hcn_2008,krb_ni_2008}. While direct laser cooling has shown progress in achieving colder and denser samples \cite{tarbutt_subdoppler,mot_doyle}, these methods still produce a factor $10^9$ lower phase-space density than coherent production of molecules from atomic mixtures.  This method has been successfully applied to several alkali mixtures and created ground state dipolar molecules \cite{krb_ni_2008,rbcs_hcn_2014,rbcs_cornish_2014,nak_park_2015,narb_wang_2016,nali_us}.

An interesting molecular species is $^{23}$Na$^6$Li.  It is a fermion with small spin-orbit coupling and a small van-der Waals length, and should therefore have a long lifetime even in the triplet ground state \cite{univrates_julienne_2011}.  Triplet NaLi ground state molecules have both an electric dipole moment (of $0.2\,\,$Debye \cite{dmoment_krems_2013}) and a magnetic moment (of $2\mu_\text{B}$), which makes them candidates to study electric dipole-dipole interactions along with magnetically tunable effects such as collisional resonances \cite{magmoment_timur_2007}.

\begin{figure}[h]
	\includegraphics[width=0.5\textwidth]{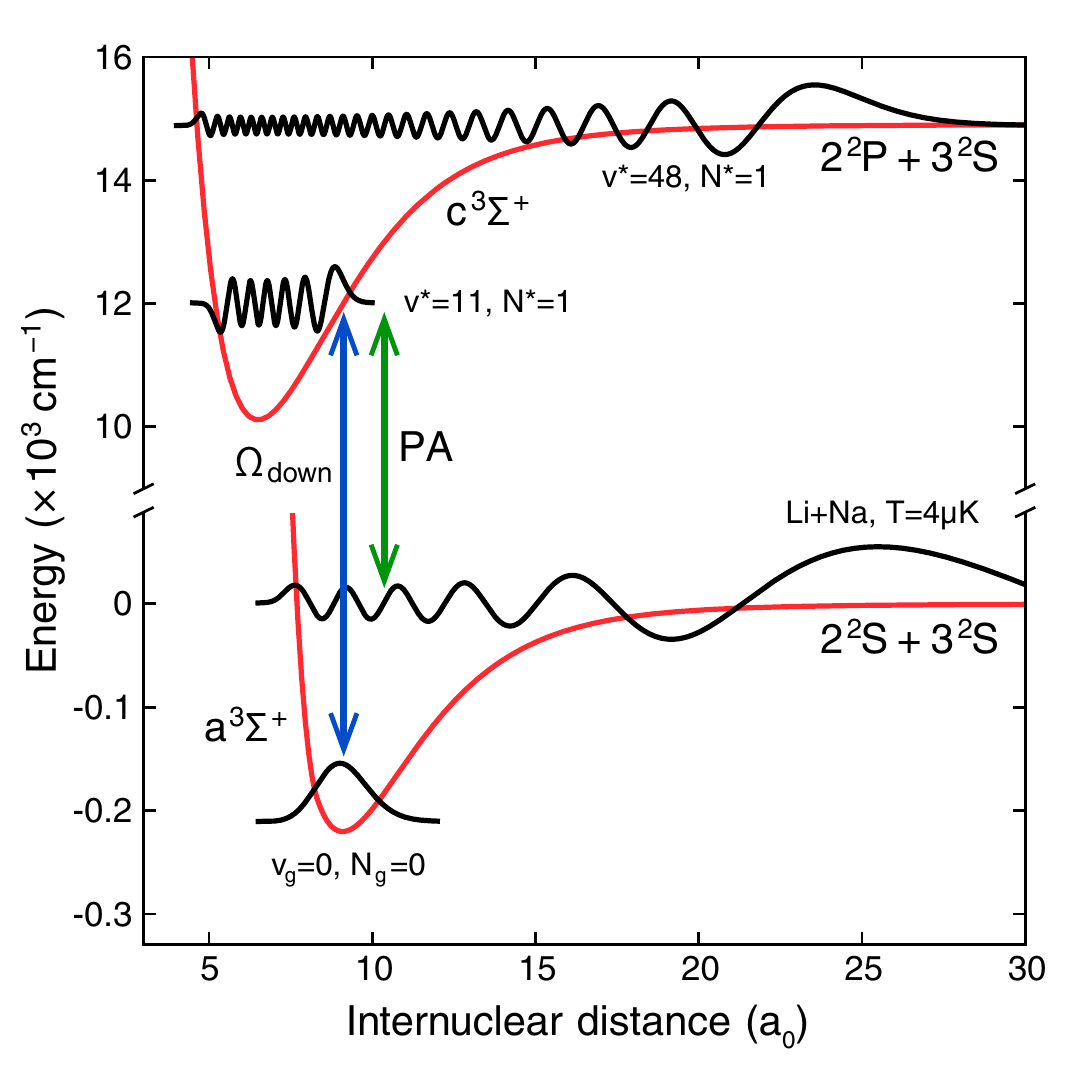}\\[-3ex]
	\caption{Energy potential diagram for the relevant NaLi triplet molecular states. A photoassociation laser is used to couple the initial scattering wavefunction to one of two excited states ($c^3\Sigma^+, v^*\,{=}\,11,48,  N^*\,{=}\,1$), which are coupled to the ground state vibrational energies with a variable frequency ``downleg'' laser used for the spectroscopic search (shown is the particular case of coupling to the ro-vibrational ground state, $a^3\Sigma^+, v_g\,{=}\,0, N_g\,{=}\,0$)}
	\label{fig:molpot}
\end{figure}

The formation of ground state molecules using stimulated Raman transitions requires the spectroscopic identification of suitable electronically excited and ground state molecular potentials for a two-photon transfer to the ro-vibrational ground state (Fig.\,\ref{fig:molpot}) \cite{nak_martin_spec,rbcs_nagerl}.
In the case of NaLi, sparse experimental data on the triplet potentials was available prior to our work: There were no prior spectroscopic observations of the excited triplet molecular states, while the ground $a^3\Sigma^+$ potential has only been studied through atomic collisional properties, particularly the observation of Feshbach resonances from highly excited vibrational states ($v_g\,{=}\,10$, where $v_g$ is the vibrational quantum number) \cite{nali_feshbach_original,feshspec_tiemann_2012}. Conventional spectroscopic efforts to observe the triplet potentials using hot NaLi mixtures were unsuccessful due to the small spin-orbit coupling in the NaLi system.  As a result, no fluorescence involving triplet states was observed after photo-excitation of an initial mixture of singlet NaLi molecules \cite{nali_singlet_tiemann}. 

In the preceding paper, we reported a spectroscopic study of excited triplet states of NaLi populated by single-photon transitions from free colliding Na and Li atoms (photoassociation) \cite{in_prep_footnote}.  In this paper, we report on the direct observation of all 11 vibrational states in the triplet ground state potential via two-photon spectroscopy, again starting from free Na and Li atoms. Using this spectroscopic information, we have recently succeeded in creating  triplet NaLi ground state molecules from an ultracold mixture of Na and Li with the well known procedure of Feshbach molecule formation \cite{feshmol_ketterle_2012} followed by stimulated Raman adiabatic passage (STIRAP) \cite{nali_us}. These are the first ultracold molecules with both electric and magnetic dipole moments.

This work fills a gap in our knowledge of bi-alkalis. For all other bi-alkalis, the ground state potential has been well-characterized \cite{lik_asigma,lirb_asigma,lics_sigma,nak_xsigma,narb_allsigma,nacs_allsigma,krb_ground_sigma,kcs_ground_sigma,rbcs_demille}. Here we report the binding energies of all vibrational states in the NaLi $a^3\Sigma^+$ potential, bound-to-bound transition strengths to the $c^3\Sigma^+$ potential, and improved potential fit parameters which build on prior \textit{ab initio} calculations \cite{nali_singlet_tiemann}.

\begin{figure*}[!ht]
	\includegraphics[width=\textwidth]{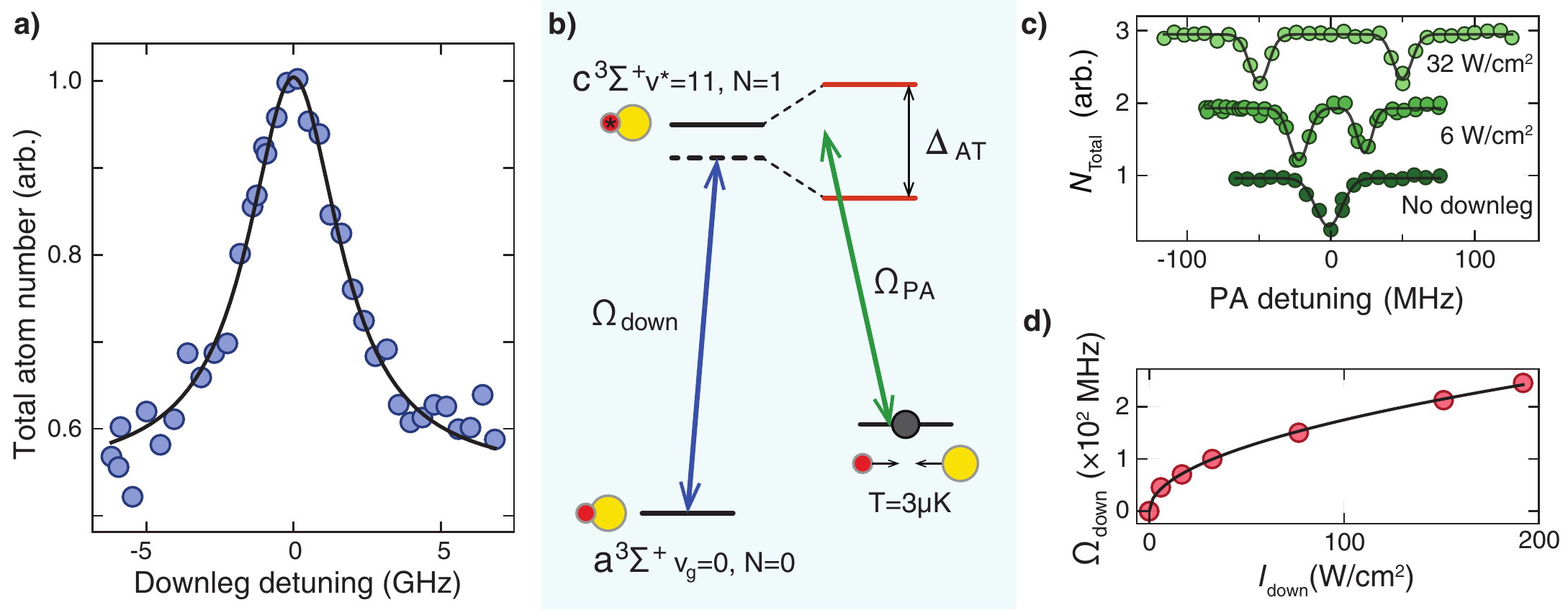}%
	\caption{Sample two-photon spectra using the $c^3\Sigma^+$, $v^*\,{=}\,11,N^*\,{=}\,1$ excited state and the  $a^3\Sigma^+$, $v_g\,{=}\,0,N_g\,{=}\,0$ ground state. a) A sample dark-resonance spectrum obtained when sweeping the downleg laser while keeping the PA laser on resonance, fit to the lineshape model described by eq.\,\eqref{eq:lineshape}. b) Energy level diagram showing the isolated three-level system used in the spectroscopic search (shown here for $v^*\,{=}\,11$, $v_g\,{=}\,0$, however the diagram is conceptually the same for other $v^*,v_g$ combinations). The downleg coupling perturbs the excited state, leading to the observation of a dark-resonance feature or an Autler-Townes doublet, depending on which laser frequency is varied. c) A sample Autler-Townes splitting obtained when sweeping the PA laser while keeping the downleg on resonance. d) The downleg Rabi frequencies obtained from the Autler-Townes data, fit to the expected scaling of the Rabi frequency $\Omega_\text{down}\,{\propto}\,\sqrt{I}$, where $I$ is the downleg laser intensity.}
	\label{fig:bigfigure}
\end{figure*}
\section{Two-photon spectroscopic search}

We produce a dual species mixture of Na in the $|F\,{=}\,2,m_F\,{=}\,2\rangle$ and Li in  $|F\,{=}\,3/2,m_F\,{=}\,3/2\rangle$ states, confined in a cigar shaped Ioffe-Pritchard magnetic trap (see \cite{in_prep_footnote}). The mixture has peak densities of $n_\text{Na}\,{=}\,9 \,{\times}\, 10^{12}\,\text{cm}^{-3}$ of Na and $n_\text{Li}\,{=}\,3 \,{\times}\, 10^{12}\,\text{cm}^{-3}$ of Li, at a temperature of $T=3.7\,\,\mu$K with degeneracy parameters $T/T_c\,{=}\,2.7$ and $T/T_F\,{=}\,1$, where $T_c$ is the Na condensation temperature and $T_F$ is the Li Fermi temperature.  The bias field of the magnetic trap is 1.2\,G aligned along the long ($z$) direction of the trap. The free atoms are detected by imaging with resonant light in absorption.

We use $v_g$, $N_g$ to refer to the vibrational and rotational quantum numbers in the ground $a^3\Sigma^+$ potential, and similarly $v^*$, $N^*$ for the electronically excited states in the $c^3\Sigma^+$ potential. The spectroscopic search for vibrational ground states $v_g$ was performed by simultaneously illuminating the ultracold mixture with two beams parallel to the magnetic field ($z$) direction: A photoassociation (PA) laser that is tuned to a known free-to-bound PA resonance with intermediate state $v^*$ (for which $N^*\,{=}\,1$ is fixed by selection rules), and a ``downleg'' laser which is swept in search of a two-photon resonance between the $s$-wave two-body collisional state and vibrational state $v_g$ (Fig.\,\ref{fig:molpot}). If the downleg frequency is not resonant with a $v^* \,{\leftrightarrow}\, v_g$ transition, loss of free atoms is observed due to resonant PA to the excited state $v^*$ followed by spontaneous decay into lower lying molecular states. However if the downleg frequency is resonant with a bound-to-bound $v^*\,{\leftrightarrow}\, v_g$ transition, the excited state $v^*$ is Stark shifted away from PA resonance, resulting in suppression of PA loss which we refer to as a ``dark-resonance'' feature as shown in Fig.\,\ref{fig:bigfigure}a,b \cite{dark_res_ref,krb_ni_2008,rbcs_nagerl,nak_martin_spec}. The center of the dark-resonance feature gives the position of the ground state $v_g$ (the exact lineshape fitting function is described in Sec. \ref{sec:linestrengths}).
 
In order to maximize the dark-resonance signal, two intermediate excited states were used ($v^* \,{=}\, 48,11$) to keep the Franck-Condon factors for the bound-to-bound $v^*\,{\leftrightarrow}\, v_g$ transition large for both weakly and deeply bound states of the $a^3\Sigma^+$ potential. We began the spectroscopic search near dissociation of the $a^3\Sigma^+$ potential, using as guidance prior observation of the $v_g\,{=}\,10$ state in Feshbach resonances \cite{nali_feshbach_original,feshmol_ketterle_2012,feshspec_tiemann_2012}, and the $a^3\Sigma^+$ \textit{ab initio} potential \cite{nali_singlet_tiemann}.  PA to a near-dissociation state of the $c^3\Sigma^+$ potential ($v^*\,{=}\,48$, $f_\text{PA}\,{=}\, 446.32951(1)$\,\,THz) was used to obtain dark-resonance spectra for the  $v_g\,{=}\,10,9$ ground states using an external cavity diode laser for the PA beam ($1$\,\,mW), and a tunable $670\,{-}\,725\,$nm Ti:Sapphire laser (50\,\,mW) for the downleg beam, both with beam waists of $200\,\,\mu$m. We did not observe the $v_g\,{=}\,8$ state using the near-dissociation intermediate state due to the sudden drop in $v_g\,{\leftrightarrow}\, v^*$ Franck-Condon factor (see Sec. \ref {sec:linestrengths}, Fig.\,\ref{fig:FC}). This necessitated the use of PA to a deeper bound excited state ($c^3\Sigma^+$, $v^*\,{=}\,11$, $f_\text{PA}\,{=}\, 359.99799(1)$\,\,THz) to find the remaining vibrational lines, for which the PA and downleg laser beams were produced by two tunable $725\,{-}\,950\,$nm Ti:Sapphire lasers ($300$\,\,mW, beam waists of 160\,\,$\mu$m). In all cases, the laser light for both the PA and downleg beams was linearly polarized, giving $\sigma^+ + \sigma^-$ polarization relative to the magnetic field. The exposure times were $200\,{-}\,500$\,\,ms. 

The vibrational state positions obtained from the spectroscopic search are given in Table \ref{table:positions}. The vibrational state binding energy was determined by taking the difference between the PA and downleg laser frequencies, hence the measurement error is independent of the uncertainty in the intermediate ($c^3\Sigma^+$) state position. The dark-resonance spectra of all vibrational states showed only one Lorentzian feature (for example, Fig.\,\ref{fig:bigfigure}a). 

\begin{table}[!htb]
	\caption{NaLi $a^3\Sigma^+$ vibrational state binding energies. The uncertainty stems from fitting dark-resonance features. In the case of $v_g\,{=}\,10,9,0$, Autler-Townes spectra (see text) were used in binding energy determination. These produce much narrower features, thus the uncertainty is limited by the wavemeter to ${\pm}\,10$\,MHz.}\label{table:positions}
	\begin{minipage}{.5\columnwidth}
		\begin{ruledtabular}\begin{tabular}{cc}
				\,\,$v_g$ & Binding Energy (GHz)\\
				10 & 9.35(1) \\
				9 & 72.36(1) \\
				8 & 232.88(4) \\
				7 & 517.13(3) \\
				6 & 935.64(3)\\
				5 & 1490.63(3) \\
		\end{tabular}\end{ruledtabular}
	\end{minipage}%
	\vline
	\begin{minipage}{.5\columnwidth}
		\begin{ruledtabular}\begin{tabular}{cc}
				\,\,$v_g$ & Binding Energy (GHz) \\
				4 & 2180.38(6) \\
				3 & 3003.00(4) \\
				2 & 3954.92(6) \\
				1 & 5033.82(3) \\
				0 & 6238.16(1) \\
				& \\
		\end{tabular}\end{ruledtabular}
	\end{minipage} 
\end{table}

The vibrational quantum number assignment of the $a^3\Sigma^+$ states was determined by comparison to \textit{ab initio} potentials. The observed position of the least-bound state with binding energy of $9.35(1)\,\,$GHz is in agreement with observations of the $v_g\,{=}\,10, N_g\,{=}\,0$ state in Feshbach resonance spectra \cite{feshspec_tiemann_2012}, and deeper bound vibrational states deviated from the expected positions based on \textit{ab initio} potentials by ${<}\,3$\,\,\% in binding energy. We have also searched the region below the $v_g\,{=}\,0$ state, and verified the absence of any observable feature.  The rotational quantum numbers of the ground states can have two possible values, $N_g\,{=}\,0,2$, from the $\Sigma \,{ \leftrightarrow}\, \Sigma$ selection rule $\Delta N\,{=}\,1$. For the ro-vibrational ground state, $v_g\,{=}\,0$, $N_g\,{=}\,0$, we confirmed the rotational quantum number by spectroscopically finding the $N_g\,{=}\,2$ state $27.7(1)$\,\,GHz above the ro-vibrational ground state, which corresponds to a rotational constant $B(v_g\,{=}\,0)\,{=}\,4.63(2)$\,\,GHz (excluding centrifugal distortion), consistent with \textit{ab initio} predictions. The $N_g\,{=}\,2$ dark-resonance spectrum showed a manifold of states spanning 200\,\,MHz, however we have not made a conclusive identification of quantum numbers for the hyperfine states in this manifold, and we report the average line position. Having established the rotational quantum numbers of the least-bound state and the ro-vibrational ground state, the rotational quantum numbers of the remaining states $0\,{<}\,v_g\,{<}\,10$ are inferred to be $N_g\,{=}\,0$ because we did not observe any discontinuous jumps in the line positions on the scale of the expected $\Delta N_g\,{=}\,2$ splitting when compared to \textit{ab initio} theory. Knowing the rotational quantum number $N_g\,{=}\,0$, we can more precisely identify the observed ground vibrational state: When using the $v^*\,{=}\,11$ intermediate state, the PA laser was resonant to the stretched state, $J\,{=}\,2, m_J\,{=}\,2$ \cite{in_prep_footnote}, thus by selection rules we conclude the dark-resonance spectra must be from coupling to the stretched ground state, $J\,{=}\,1, m_J\,{=}\,1$.

\section{Line strengths}\label{sec:linestrengths}
\begin{figure}
	\includegraphics[width=0.5\textwidth]{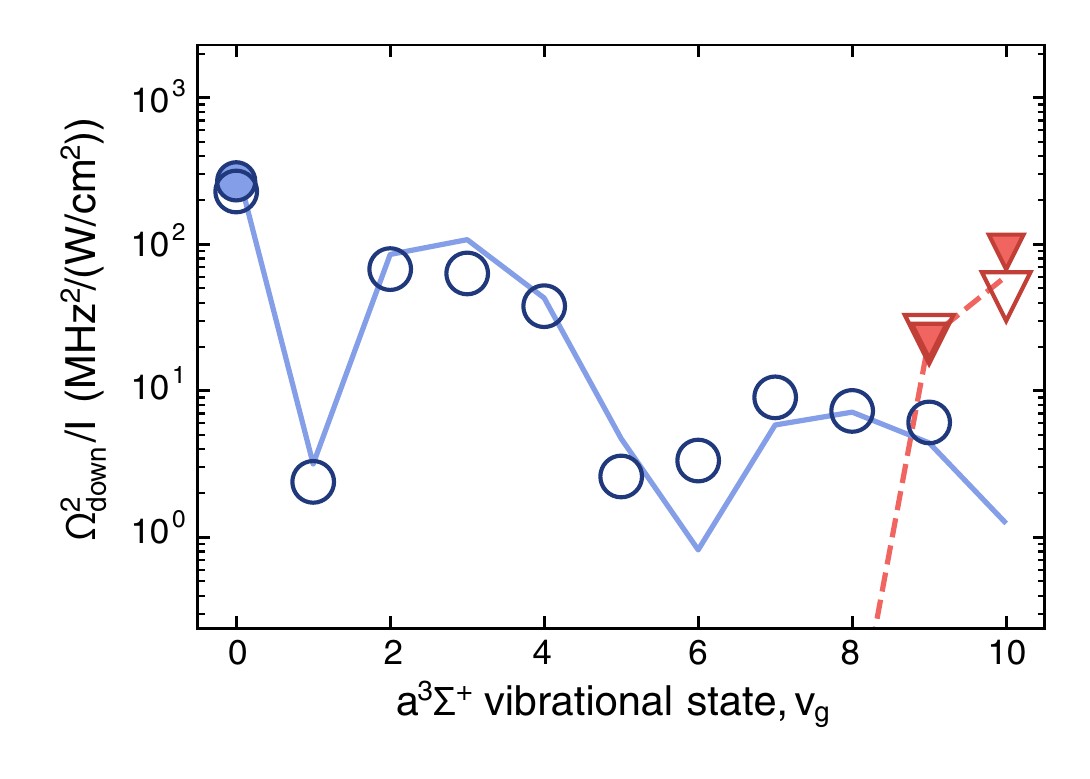}
	\caption{Rabi frequencies for bound-to-bound (downleg) transitions between vibrational states in the ground $a^3\Sigma^+$ potential and the excited $c^3\Sigma^+$ potential. Blue circles represent measured Rabi frequencies to the $v^*\,{=}\,11$ excited state; red triangles for excitations to $v^*\,{=}\,48$. Solid color data points are Rabi frequencies measured from Autler-Townes spectra (ie. the observed line splitting), while hollow points used eq.\,\eqref{eq:lineshape} to extract the downleg Rabi frequency from dark-resonance spectra.  The solid and dashed lines are theoretical predictions from evaluation of Franck-Condon overlap integrals in \textit{ab initio} potentials (Note: The theory curves are shown as continuous lines for visual clarity; they do not have physical meaning for non-integer $v_g$.)}\label{fig:FC}
\end{figure}

The line strengths of the downleg $v^* \,{\leftrightarrow}\, v_g$ transitions can be determined from spectra in which one of the lasers is swept, while the other is held on resonance to the excited state $v^*$. The symmetry of the three-level system is broken by the choice of free atoms as the initial state, thus the lineshapes from sweeping either the PA or downleg laser will be qualitatively different (Fig.\,\ref{fig:bigfigure}b). In the case of sweeping the PA laser while keeping the downleg on resonance, the excited state is dressed with the ground state to form an Autler-Townes doublet (Fig.\,\ref{fig:bigfigure}b,c), and the strength of the downleg Rabi frequency ($\Omega_\text{down}$) is given simply by the observed line splitting. We have performed Autler-Townes spectroscopy using the $v_g\,{=}\,0 \,{\leftrightarrow}\, v^*\,{=}\,11$ and $v_g\,{=}\,9,10\,{\leftrightarrow}\,v^*\,{=}\,48$ states to determine the downleg Rabi couplings (Fig.\,\ref{fig:FC}, solid points). However, our primary experimental effort was to find the previously unknown binding energies of the ground states $v_g$, hence the majority of data was taken with the opposite, dark-resonance configuration, where the downleg laser was swept with the PA on resonance.  In this case, the dark-resonance feature is caused by an AC Stark detuning of PA to the intermediate state $v^*$, thus the relative strength of the bound-to-bound couplings can be determined from the lineshape widths. In the limit where the PA is perturbative, the atom number lost from the PA follows a Lorentzian, with the detuning determined by the downleg AC Stark shift. The remaining atom number has the form (see appendix):
\begin{equation}\label{eq:lineshape}
\dfrac{N}{N_0}\sim 1- \dfrac{1}{1+\left(\delta_\text{AC}/\left(\Gamma/2\right)\right)^2}\,, \quad \delta_\text{AC}=\dfrac{\Omega_\text{down}^2}{4\Delta_\text{down}}
\end{equation}
where $N, N_0$ is the remaining and initial atom number, respectively, $\delta_\text{AC}$ is the AC Stark shift due to the downleg laser, $\Delta_\text{down}$ is the downleg detuning, $\Omega_\text{down}$ is the downleg Rabi frequency, and $\Gamma \,{=}\, 2\pi \,{\times}\, 9(3)\,\,$MHz is the excited state linewidth, which we estimate from our prior PA spectroscopy \cite{in_prep_footnote}. Fig.\,\ref{fig:bigfigure}a shows a sample spectrum fit to eq.\,\eqref{eq:lineshape} (along with a constant vertical  offset), from which we can determine the downleg Rabi coupling $\Omega_\text{down}$. We apply this procedure to all observed ground state vibrational levels to measure the bound-to-bound ($v_g \,{\leftrightarrow}\, v^*=11,48$) Rabi couplings as shown in Fig.\,\ref{fig:FC}. The downleg Rabi frequencies measured in this way are in agreement with the values obtained from Autler-Townes spectra.  Throughout all measurements the PA Rabi frequency was ${\leq}\,10$\,kHz (estimated from PA loss measurements), which is expected from the weak nature of free-to-bound transitions.

The downleg Rabi frequency measurements are fit to Franck-Condon factors obtained from numerically solving the \textit{ab initio} potentials, and the only fitting parameter is an overall scaling factor (Fig.\,\ref{fig:FC}). The \textit{ab initio} results are in good agreement with the data, and show the importance of using the $v^*\,{=}\,11$ state to find the ro-vibrational ground state $v_g\,{=}\,0$, as the $v^*\,{=}\,48$ intermediate state shows a sharp drop in Franck-Condon factor to lower vibrational ground states, $v_g\,{\leq}\,8$. In our work on creating triplet ground state molecules, the $v^*\,{=}\,11$ state was chosen to optimize the coupling to the ro-vibrational ground state.  Spontaneous decay from $v^*\,{=}\,11$ has a ${\sim}\,50\,\,\%$ branching ratio into $v_g\,{=}\,0$ (ignoring any rotational selection rule factors and singlet-triplet transitions). Such a large Franck-Condon factor is typical of the triplet NaLi system due to the light mass of the molecule, resulting in fewer vibrational states in comparison to heavier species. This favorable branching ratio could enhance previously implemented molecule formation schemes relying on spontaneous emission \cite{pa_formation_lics,rbcs_demille,pa_formations_rbcs}, or provide a strong line for direct molecular imaging on a non-cycling transition \cite{krb_imaging}. 

\section{Potential Fits}

The ground state vibrational binding energies were fit to an $X$-representation potential model, which is a piece-wise parametrization containing a combination of physical constants (such as dispersion coefficients, equilibrium inter-nuclear distance, etc.) and phenomenological parameters \cite{nali_singlet_tiemann}. The fitting procedure was as follows: First, a Rydberg-Klein-Rees (RKR) potential was constructed using the measured binding energies and calculated rotational constants from an \textit{ab initio} potential \cite{olivierdulieu}. The point-wise RKR potential was then improved to more closely match the measured binding energies. The initial $X$-representation parameters were obtained by minimizing the root-mean-square (RMS) deviation for each point of this point-wise potential using the Broyden-Fletcher-Goldfarb-Shanno (BFGS) algorithm. The $X$-representation in the intermediate range $(R_{\rm i} < r < R_{\rm o}$, see Table\,\ref{table:fitparams}\,) is a power-series expansion. The number of expansion coefficients, $a_i$, was increased until the fit did not converge with better precision. Their values were determined by iteratively fitting outward from the bottom of the potential. 

This $X$-representation fit was further optimized through simulated annealing using the objective function
\begin{equation}
f=\sqrt{\sum^N_{k=1}\left(\dfrac{E_k^\text{obs}-E_k^\text{calc}}{\sigma_k}
\right)^2},
\end{equation}
 where $N\,{=}\,11$ is the number of measured values and $E_k^\text{obs}$ and $E_k^\text{calc}$ are the observed and calculated binding energies, respectively. The weight $\sigma_k$ is the uncertainty of our binding energy measurement (see Table\,\ref{table:positions}\,), with the exception of $v_g\,{=}\,10$ for which we use the more precise result obtained from Feshbach spectroscopy \cite{feshspec_tiemann_2012}. The free parameters used in the fitting process are denoted with asterisk signs in Table\,\ref{table:fitparams}. Since binding energies calculated from the initial $X$-representation fit of the point-wise potential already agreed with the measurements by ${\lesssim}\,1$\% on average, our goal was to improve this prior parametrization rather than find a minimal set of parameters.

The results of the simulated annealing are reported in Table\,\ref{table:fitparams}. 
\begin{table}[!htb]
	\caption{NaLi $a^3\Sigma^+$ $X$-representation potential fit parameters. Also given are the functional forms of the $X$-representation, which is segmented into an inner, middle, and outer region (for details, see \cite{nali_singlet_tiemann}).  Parameters marked with $(*)$ were varied during the fitting procedure.}
	\begin{ruledtabular}\begin{tabular}{lc}\label{table:fitparams}
			Parameter & Value \\
			\hline\hline
			\multicolumn{2}{c}{$r<R_i=3.95877\,\,\si{\angstrom}$} \\ 
			\multicolumn{2}{c}{$V(r)=A+\dfrac{B}{r^{N_s}}$} \\
			$A\ (\text{cm}^{-1})\ (*)$ & $-3.61655 \times 10^{2}$ \\
			$B\ (\text{cm}^{-1}\si{\angstrom}^{N_s})\ (*)$ & $1.04254 \times 10^{6}$  \\
			$N_s$ & $6$  \\
			\hline
			\multicolumn{2}{c}{$R_i \leq r \leq R_o=10.6195\,\,\si{\angstrom}$} \\ 
			\multicolumn{2}{c}{$V(X)=\displaystyle\sum^6_{i=0} a_i X^i,\,\, X=\dfrac{r-R_m}{r+bR_m}$}\\
			$R_m\ (\si{\angstrom})\ (*)$ & $4.70154$ \\
			$b\ (*)$ & $7.44923\times 10^{-1}$\\
			$a_0\ (\text{cm}^{-1})\ (*)$ & $-2.29753 \times 10^{2}$  \\
			$a_1\ (\text{cm}^{-1})\ (*)$ & $-3.92982 \times 10^{-5}$  \\
			$a_2\ (\text{cm}^{-1})\ (*)$ & $9.37416 \times 10^{3}$  \\
			$a_3\ (\text{cm}^{-1})\ (*)$ & $-3.98395 \times 10^{4}$  \\
			$a_4\ (\text{cm}^{-1})\ (*)$ & $6.48866 \times 10^{4}$  \\
			$a_5\ (\text{cm}^{-1})\ (*)$ & $-3.72622 \times 10^{4}$  \\
			$a_6\ (\text{cm}^{-1})\ (*)$ & $-1.82264 \times 10^{3}$  \\
			
			\hline
			\multicolumn{2}{c}{$R_o < r$} \\
			\multicolumn{2}{c}{ $V(r)=-\dfrac{C_6}{r^6}-\dfrac{C_8}{r^8}-\dfrac{C_{10}}{r^{10}}-E_\text{ex},\,\, E_\text{ex}=B_\text{ex}r^\alpha e^{-\beta r}$}\\
			$C_6\ (\text{cm}^{-1}\si{\angstrom}^{6})\ (*)$ & $6.88556 \times 10^{6}$  \\
			$C_8\ (\text{cm}^{-1}\si{\angstrom}^{8})\ (*)$ & $1.30886 \times 10^{8}$  \\
			$C_{10}\ (\text{cm}^{-1}\si{\angstrom}^{10})\ (*)$ & $2.96200 \times 10^{9}$  \\ 
			$B_\text{ex} \ (\text{cm}^{-1}\si{\angstrom}^{-\alpha})$ & $1.80339 \times 10^{3}$ \cite{nali_singlet_tiemann}\\ 
			$\alpha$ & $4.62749$ \cite{nali_singlet_tiemann}  \\
			$\beta\ (\si{\angstrom}^{-1})$ & $2.35101$ \cite{nali_singlet_tiemann} 		
	\end{tabular}\end{ruledtabular}
\end{table}
The vibrational binding energies calculated using this improved $X$-representation give an RMS error from the measured values of $700\,{\rm MHz}$. This is a factor of $60$ improvement in RMS error and a factor of $100$ improvement in the objective function when compared to the potential extrapolated from observations of only the highest vibrational bound state using Feshbach spectroscopy \cite{nali_singlet_tiemann}. Our spectroscopic measurements are of the stretched hyperfine state in each vibrational level, and since the hyperfine splittings are expected to be ${\sim}\,1$\,\,GHz \cite{nali_us}, we did not attempt to optimize the potential fit any further.  The fitting process modified the leading van der Waals coefficient, $C_6$, by -2.6\% compared to its theoretical value \cite{c6}. While well outside the theoretical uncertainty, this value is fairly close to the value obtained in \cite{nali_singlet_tiemann}. The higher dispersion coefficients $C_8$ and $C_{10}$ shifted by -1.8\% and -14\% from their theoretical values \cite{c8c10}, respectively, placing them near the theoretical uncertainties. We observed that by slightly increasing $R_{\rm o}$, the fit could be optimized keeping all the dispersion coefficients within their theoretical uncertainty limits. However, this came at the expense of much worse continuity and/or differentiability of the potential across the boundary, $R_{\rm o}$. Additional information, such as knowledge of more rotational states of the high vibrational levels could help in optimizing the long-range parameters without sacrificing continuity and differentiability. 

The contribution of the exchange energy terms ($B_{ex}, \alpha,$ and $\beta$, see Table\, \ref{table:fitparams}) to the potential energy and the continuity/differentiability at the long range was less significant than those of the dispersion terms, so we used the previously reported values \cite{nali_singlet_tiemann}. An interesting observation is that the optimal point-wise potential performed better, with an RMS error of $300$\, MHz. This was probably because the point-wise potential was less constrained and yet always continuous and differentiable due to the use of cubic splines to smooth between relevant points. Nonetheless, we preferred the $X$-representation because of its physically meaningful constants (e.g., $a_0$ is the well depth).  

 Another approach to determining dispersion coefficients is the use of a LeRoy-Bernstein expansion valid for states near dissociation, which we have used in the preceding work on the NaLi $c^3\Sigma^+$ potential \cite{in_prep_footnote}. In contrast, such an approach is not possible for the $a^3\Sigma^+$ potential due to the small $C_6$ van der Waals coefficient. This makes all but the highest vibrational state wavefunction fit within the LeRoy radius, a rough inner bound for where a LeRoy-Bernstein analysis is valid \cite{leroy_radius}. 

\section{Conclusions}

We have directly observed all vibrational states in the NaLi $a^3\Sigma^+$ ground state potential using two-photon spectroscopy. Vibrational binding energies and bound-to-bound transition strengths to states in the $c^3\Sigma^+$ have been measured and are in agreement with \textit{ab initio} calculations \cite{nali_singlet_tiemann}. These observations fill a long standing hole in bi-alkali molecular spectroscopy, in which NaLi was the only molecule with missing triplet ground state photoemission spectra. This spectroscopic data was used recently in formation of ultracold triplet ground state NaLi molecules, and constitutes the foundation for manipulation of these molecules in future experiments \cite{nali_us}. 

\section*{Conflicts of interest}
There are no conflicts to declare. 

\begin{acknowledgments}
We would like to thank Li Jing and Yijun Jiang for experimental assistance. We acknowledge support from the NSF through the Center for Ultracold Atoms and award 1506369, from the ONR DURIP grant N000141613141, and from MURIs on Ultracold Molecules (AFOSR grant FA9550-09-1-0588) and on Quantized Chemical Reactions of Ultracold Molecules (ARO grant W911NF-12-1-0476). H.S. and J.J.P. acknowledge additional support from the Samsung Scholarship. 
\end{acknowledgments}

\appendix*

\section{Dark-Resonance Lineshape}
 In determining the dark-resonance lineshape, we use two assumptions: (i) the PA strength is weak ($\Omega_\text{PA}\ll\Gamma$) allowing us to ignore effects of PA saturation \cite{bohn_julienne_PAsat} and (ii) the atom loss is small, resulting in a constant PA loss rate determined by the initial atomic density.

The dark-resonance feature occurs due to a suppression of PA caused by an AC Stark shift of the excited state $v^*$ due to the downleg coupling. In the absence of a downleg laser, the effect of PA to the state $v^*$ is described by
\begin{equation}\label{eq:parate}
\dot n =-Kn^2 \> \rightarrow \> n(\tau)=\dfrac{n_0}{1+n_0 K \tau}\approx n_0(1-n_0K\tau)
\end{equation}
where $n,n_0$ is the final and initial atom number density, which is related to the total atom numbers $N,N_0$ through the effective volume, $V_\text{eff}=N^2/\int n^2 dV$. $\tau$ is the PA exposure time, $K$ is the two-body loss coefficient, and the last step makes the assumption that the PA loss is small and can be modeled with a constant rate depending only on the initial density $n_0$. The PA loss coefficient is given by \cite{bohn_julienne_PAsat}:
\begin{equation}\label{eq:pasat}
K \propto \dfrac{\Gamma \Omega_\text{PA}}{\Delta_\text{PA}^2+\left[\left(\Omega_\text{PA} + \Gamma\right)/2\right]^2} \propto \dfrac{1}{\Delta_\text{PA}^2+\left(\Gamma/2\right)^2}
\end{equation}
where $\Omega_\text{PA}$, $\Delta_\text{PA}$ is the PA Rabi frequency and detuning, respectively, $\Gamma$ is the excited state linewidth, and the PA Rabi frequency is well below saturation ($\Omega_\text{PA}\ll\Gamma$). During the spectroscopic search the PA is resonant, ie. $\Delta_\text{PA}\,{=}\,0$, however the presence of the downleg shifts the intermediate state, effectively detuning the PA from resonance by
\begin{equation}\label{eq:twolevelshift}
\Delta_\text{PA}=\frac{1}{2}\left(\sqrt{\Delta_\text{down}^2 + \Omega_\text{down}^2}-\Delta_\text{down}\right)\\
\approx \dfrac{\Omega_\text{down}^2}{4\Delta_\text{down}}
\end{equation}
where $\Omega_\text{down}$, $\Delta_\text{down}$ are the downleg Rabi frequency and detuning, respectively, and we have assumed large detunings, $\Delta_\text{down}\,{\gg}\,\Omega_\text{down}$. (Note, eq.\,\eqref{eq:twolevelshift} is the typical result from a two-level system formed by the ground and excited molecular states). By substitution of eq.\,\eqref{eq:twolevelshift}, \eqref{eq:pasat} into eq.\,\eqref{eq:parate}, we reach the final fitting function which we use to measure $\Omega_\text{down}$: 
\begin{equation}\label{eq:fitshape}
\dfrac{N}{N_0}= A\left(1-\dfrac{1}{1+\left(\Omega_\text{down}^2/(2\Delta_\text{down}\Gamma)\right)^2}\right) + B
\end{equation}
where $A,B$ are fitting parameters.

\bibliographystyle{apsrev4-1}

%

\end{document}